# Increasing power and robustness in screening trials by testing stored specimens in the control arm

Hormuzd A. Katki[1,*], Li C. Cheung[1]

[1]Division of Cancer Epidemiology and Genetics, National Cancer Institute, National Institutes of Health, Department of Health and Human Services, Bethesda, Maryland

*** Corresponding author:**
**HAK:** Division of Cancer Epidemiology and Genetics, National Cancer Institute, 9609 Medical Center Dr., Room 7E592, Bethesda, MD 20892, Phone: 240-276-7423, katkih@mail.nih.gov

**Word Count**: 2886/3000

**Funding/Support/IRB statement:** This study was supported by the US National Institutes of Health.

**Role of the Funding source:** The NIH had no role in the design and conduct of the study; in the collection, analysis, and interpretation of the data; or in the preparation, review, or approval of the manuscript. The authors alone are responsible for the views expressed in this article and they do not necessarily represent the views, decisions, or policies of the institutions with which they are affiliated.

**Conflicts of Interest**: None

**Acknowledgements:** We thank Philip Prorok and Paul Pinsky (NCI) for helpful comments on earlier versions of this manuscript.

**Abstract** (280/350 words)

**Background:** Screening trials require large sample sizes and long time-horizons to demonstrate mortality reductions. We recently proposed increasing statistical power by testing stored control-arm specimens, called the "Intended Effect" (IE) design. To evaluate feasibility of the IE design, the US National Cancer Institute (NCI) is collecting blood specimens in the control-arm of the NCI Vanguard Multicancer Detection pilot feasibility trial. However, key assumptions of the IE design require more investigation and relaxation.

**Methods:** We relax the IE design to (1) reduce costs by testing only a stratified sample of control-arm specimens by incorporating inverse-probability sampling weights, (2) correct for potential loss-of-signal in stored control-arm specimens, and (3) correct for non-compliance with control-arm specimen collections. We also examine sensitivity to "unintended effects" of screening (i.e. outcome rates in screen-negatives increasing due to false reassurance).

**Results:** In simulations, testing all primary-outcome control-arm specimens and a 50% sample of the rest maintains nearly all the power of the IE while only testing half the control-arm specimens. Power remains increased from the IE analysis (versus the standard analysis) even if unintended effects exist. The IE design is robust to some loss-of-signal scenarios, but otherwise requires retest-positive fractions that correct bias at a small loss of power. The IE can be biased and lose power under control-arm non-compliance scenarios, but corrections correct bias and can increase power.

**Conclusions:** The IE design can be made more cost-efficient and robust to loss-of-signal. Unintended effects will not typically reduce the power gain over the standard trial design. Non-compliance with control-arm specimen collections can cause bias and loss of power that can be mitigated by corrections. Although promising, practical experience with the IE design in screening trials is necessary.

## Introduction

Randomized controlled trials (RCTs) for screening tests require large sample sizes and long time-horizons to demonstrate mortality reductions.[1–3] For example, the NHS-Galleri Multicancer Detection (MCD) screening trial randomized 140,000 people and requires 8-10 years to examine cancer mortality results.[4] MCD tests use a single blood specimen to screen for cancer at multiple organ sites simultaneously, including those without recommended screening, which is an exciting prospect.[5,6] However, RCT data on the clinical benefits and harms of MCD-based screening will be unavailable for many years. Yet, in the US, MCD tests are already being sold, underscoring the need to provide RCT evidence as soon as possible.[7]

To increase statistical power in screening trials, we[8] and others[9–11] have proposed testing stored material among controls and comparing trial outcomes only among those who ever test positive, which we call the “Intended Effect” (IE) design[8]. The *intended effect* of screening is that only those who test-positive at least once (“ever-positive”) could receive treatments which affect mortality. In contrast, those who never test-positive (“never-positive”) never undergo diagnostic procedures (much less treatment) and thus their mortality should be unaffected by screening. Thus, removing everyone who likely could not be affected by screening (i.e. removing never-positives), and comparing outcomes only among ever-positives in both arms, can have substantially greater statistical power.[8] An MCD RCT relying the IE design could halve the sample-size required to attain 90% power.[8]

To evaluate feasibility of the IE design, the US National Cancer Institute (NCI) is collecting blood specimens in the control-arm of the NCI Vanguard MCD pilot feasibility RCT.[12] Although promising, the IE design requires assumptions that have not been adequately investigated or relaxed. We relax the IE design to (1) reduce costs by testing only a stratified sampling of control-arm specimens, (2) account for potential loss-of-signal in stored control-arm specimens, and (3) correct for non-compliance with control-arm specimen collections. We also examine sensitivity to “unintended effects” of screening (i.e. outcome rates in screen-negatives increasing due to false reassurance).

## Methods

### *Brief review of the Intended Effect (IE) Design/Analysis*

**Figure 1** shows a schematic of the IE analysis for an example idealized MCD trial. The standard trial analysis, with 50k individuals per-arm, 2% cancer death rate, and a screening relative risk (RR) of 0.90 would have only 65% power. The IE design tests control-arm specimens and removes everyone who are never test-positive, because they should not be affected by screening as they receive no medical interventions (i.e., assumed to have “no unintended effects”), and in **Figure 1** their relative risk (”$RR_{neg}$”) equals 1. The IE analysis focuses on the ever-positives, the only people who might receive medical interventions and thus benefit from screening. The RR among ever-positives is $RR_{pos}$=0.867, which is stronger than RR=0.90. Importantly, the p-value for $RR_{pos}$ is 10-fold smaller than for the RR. To achieve 90% power at $\alpha$=0.05, the standard trial would require 98k individuals per-arm versus 53k per arm for the IE design. Hence the IE analysis, which focuses on $RR_{pos}$ and its p-value, achieves 90% power at about half the sample-size required by the standard analysis.

Secondarily, the IE analysis also calculates the risk difference and relative risk among never-positives ($RD_{neg}$ and $RR_{neg}$, respectively) to test the "no unintended effects" assumption, which is that $RD_{neg}$ =0 and $RR_{neg}$ =1.

See **Supplement** for more review and notation, and discussion of ethical issues[13,14].

*Extending the IE Analysis to only test a stratified sample of control-arm specimens*

Rather than testing all control-arm specimens, it is more cost-efficient to test everyone who experiences the primary outcome ("cases") and a stratified random sample of the others ("controls") for testing. The **Supplement** shows how to use observed sampling fractions as inverse-probability weights (IPW) to estimate the $RR_{pos}$. Stratification can be on any covariates collected on everyone in the control-arm to oversample people associated with testing ever-positive. The more predictive the covariates are for ever-positivity, the fewer controls would require testing.

*Sensitivity of IE statistical power to unintended effects of screening*

The IE design assumes that individuals who never test positive receive no medical interventions and are unaffected by screening. The "no unintended effects" assumption is analogous to the "exclusion restriction" in causal inference, that "never-takers" in both arms should experience the same outcome rates[15]. However, there may be unintended effects of screening, such as false-reassurance in the screen-arm never-positives, who may skip standard of care (SOC) screens or otherwise alter their behavior to be riskier ($RD_{neg}<0$, or equivalently, $RR_{neg}>1$), and thus a negative screen inadvertently causes harm. The opposite of false-reassurance is 'non-assurance', which could happen if people in the screen arm who test never-positive are not reassured and seek out extra screening ($RD_{neg}>0$, or equivalently, $RR_{neg}<1$).

Power is determined by $\alpha$ and the non-centrality parameter (see page 91 in[16]) The **Supplement** calculates the ratio increase in the non-centrality parameter for $RR_{pos}$ versus the standard RD, for any value of $RD_{neg}$:

$$Z_{ratio} = \left\{1 - \frac{RD_{neg}}{RD} \cdot P(M-)\right\} \cdot \sqrt{\frac{P(M+)}{P(M+|D+)P(M+|D-)}},$$

where "RD" is the overall trial risk-difference, "P(M+)" is the probability of being ever-positive (i.e. testing positive on any screen), "P(M-)" is the probability of testing never-positive, and "D+" denotes experiencing the trial outcome (e.g. cancer death). The radical is >1 because its numerator is a weighted average of the 2 factors in the denominator.  Thus $Z_{ratio} > 1$ and the IE analysis has more power than the standard analysis if there are no unintended effects ($RD_{neg}$=0) or if there is false-reassurance in the screen-arm ($RD_{neg}$<0).  If there were 'non-assurance' ($RD_{neg}$>0), the IE analysis can still have more power than the standard analysis, although not guaranteed.

**Supplement** explains how we simulated to calculate power for various scenarios.

*Correcting for loss-of-signal in stored control-arm specimens*

Stored specimens degrade over time.  There may be less positivity by testing stored specimens (in the control arm) versus testing fresh specimens (in the screen arm), which may bias $RR_{pos}$ or $RR_{neg}$.  Degradation of stored specimens rarely (if ever) results in gain-of-signal, and we assume it is impossible.  The **Supplement** demonstrates:

- If loss-of-signal is only differential by arm (i.e. non-differential by outcome), there is no bias in $RR_{pos}$ by substituting the observed ever-positivity as the true ever-positivity, although there is loss of power.  An example of loss-of-signal only differential by arm is if the stored specimens that lose signal were a simple random sample of all ever-positives.  If instead false positives were more likely to have a weak degradable signal than true positives, then loss-of-signal is differential by outcome and there is bias in $RR_{pos}$.
- For $RR_{neg}$, even if loss-of-signal is only differential by arm, there is bias by substituting the observed ever-positivity for true ever-positivity.

To correct bias, we propose also storing a portion of the specimen in the screen-arm (screen-arm conducts testing on fresh specimens) and retest all stored specimens that tested positive on the fresh specimen. Define $\widetilde{M}$ as the observed ever-positivity indicator, which degrades the true M by loss-of-signal. For simplicity, assume no gain-of-signal is possible. We estimate (1) the fraction of the screen-arm true-positives that also retest-positive ($P_1(\widetilde{M}+|M+,D+)$), and (2) the fraction of the screen-arm false-positives that also retest-positive ($P_1(\widetilde{M}+|M+,D-)$). These fractions account for the possibility that there may be more loss of signal in false-positives, which may tend to be weaker signals than true-positives. We substitute retest-positivity to unbiasedly calculate the outcome rate in control-arm ever-positives:

$$P_0(D+|M+) = \left\{1 + \frac{P_0(D-)}{P_0(D+)} \times \frac{P_0(\widetilde{M}+|D-)/P_1(\widetilde{M}+|M+,D-)}{P_0(\widetilde{M}+|D+)/P_1(\widetilde{M}+|M+,D+)}\right\}^{-1},$$

and thereby correct bias in $RR_{pos}$. See **Supplement** for derivation and demonstration of bias correction. A similar calculation to correct bias in $RR_{neg}$ is in the **Supplement**.

*Correcting for non-compliance with specimen collections in the control-arm*

We focus on non-compliance in the control-arm from skipping specimen collections (we assume the screening test cannot be obtained outside of the trial, which is a problem for the standard analysis as well). Hence non-compliance can only reduce ever-positivity. Note that, those who skipped screens but tested positive at any screen they attended, are known to be ever-positive. Thus we examine robustness to including those who both skipped screens and never tested positive at any screen they attended (i.e. those with unknown ever-positivity), into the never-positive table. In this framework, non-compliance to specimen collections is mathematically equivalent to a particularly acute loss-of-signal. Thus, the results from robustness to loss-of-signal section carry-over. Thus, just as loss-of-signal that is only differential by arm (i.e. non-differential by outcome) does not cause bias, similarly non-compliance that is only differential by arm does not cause bias (although there is loss of power). Similarly, just as loss-of-signal that is differential by outcome causes bias, non-compliance that is differential by outcome also causes bias. See **Supplement** for more discussion.

Correcting for control-arm non-compliance requires placing everyone in both arms with unknown ever-positivity into a third 2x2 table (see **Supplement)**. We do not correct estimates to perfect compliance in both arms, because perfect compliance is unrealistic. We propose that the non-compliance correction be to set the non-compliance pattern in the control-arm to equal that in the screen-arm. That is, we propose to not correct for non-compliance in the screen-arm. This is because it is appropriate to “penalize” screening if the screening test is difficult for participants to adhere to.

Non-compliance that is completely non-differential cannot cause bias. We conducted simulations to quantify bias and power in various control-arm non-compliance scenarios.

## Results

*Efficient stratified sampling of stored control-arm specimens to test*

**Table 1** simulated sampling of control-arm specimens with binomial sampling among those who develop a trial outcome and a single fraction of those who do not develop trial outcome. When sampling all controls for testing provides 89-90% power, sampling half of the controls for testing provides 84-86% power (**Table 1A**), nearly the same power while obviating testing for half the control-arm. If only 80% of cases could be tested (**Table 1B**), power drops from 89-90% to 87-88% when sampling 70% of the controls along with the 80% of cases, showing that small losses in testing of cases could be made up by testing more controls. The simulations do not include sampling strata predictive of ever-positivity, such as cancer-status, age, or other risk-factors, which in principle could make the sampling plan more efficient.

*Unintended effects and statistical power in the IE analysis*

The **Supplement** demonstrates that the IE analysis has more power than the standard analysis if there are no unintended effects ($RD_{neg}$=0, or equivalently, $RR_{neg}$=1) or if there is false-reassurance in the screen-arm ($RD_{neg}$<0). But even if there were 'non-assurance' ($RD_{neg}$>0), the IE analysis can still have more power than the standard analysis. **Figure 2** plots power versus $RR_{pos}$ for an example MCD trial with RR=0.9, as ever-positivity and $RR_{neg}$=1vary. The standard analysis has only 36% power, but many

IE analyses achieve >90% power as $RR_{pos}$ becomes smaller and as ever-positivity decreases. As expected, the power gain is greatest for false-reassurance (RRneg=1.05), then no intended effects ($RR_{neg}$=1), and least gain for non-assurance (RRneg=0.95).

The power curves are monotonic and similar for ever-positivity > 5%, and power is close to that of the standard analysis when $RR_{pos}$ =RR. However, for very small 2.5%-5% ever-positivity and $RR_{pos} \approx$ RR, power is not monotonic at $RR_{pos} \approx$ RR because the larger number of outcomes at $RR_{pos} \approx$ RR can outweigh a stronger $RR_{pos}$ <RR (see **Supplement Figure S1** for an example).

*Correcting for potential loss-of-signal from testing stored versus fresh specimens*

When loss-of-signal is non-differential by outcome, $RR_{pos}$ is unbiased with little loss in power at small loss-of-signal fractions of 10%, but substantial loss of power if 30% or more of signal were lost (**Table 2A**). However, $RR_{neg}$=1 is strongly biased even at small loss-of-signal with highly inflated type-1 error $\alpha$ (**Table 2A)**. When loss-of-signal is differential by outcome (**Table 2B)**, bias generally ensues. With a 20% loss-of-signal in non-events (and 10% in events), $RR_{pos}$ is substantially underestimated (0.867 to 0.80, a non-conservative bias).

**Table 2C** shows that our proposed corrections to $RR_{pos}$ and $RR_{neg}$=1 by using stored samples in the screen-arm are unbiased and testing for $RR_{neg}$=1 (which is null) has proper type-1 error $\alpha$. However, $RR_{pos}$ estimation inevitably loses power, e.g. if loss-of-signal is 10% in events and 20% in non-events, then power drops from 0.84 to 0.79.

*Correcting for non-compliance to specimen collections in the control-arm*

**Figure S2** shows an example of correcting for arm-differential non-compliance, where the observed and corrected $RR_{pos}$ equal that under perfect compliance. **Figure S3** shows an example of correcting for non-compliance that is differential by both arm and outcome, and the corrected $RR_{pos}$ no longer equals that under perfect compliance. This is because outcome-differential non-compliance implies that ever-positivity will be imbalanced between arms (except at the null of RR=1, $RR_{pos}$ =1, and $RR_{neg}$=1). See **Supplement** for more explanation about the figures.

**Table 3** shows simulation results from various non-compliance scenarios. Neither $RR_{pos}$ nor $RR_{neg}$ has bias under non-compliance that is non-differential, or differential by arm-only. However, power for $RR_{pos}$ is reduced from 88% to 74-77%. The non-compliance correction to $RR_{pos}$ increases power to 82-85%, showing that even if the correction is not needed to reduce bias, it can reduce variance and increase power.

However, when non-compliance is differential by both arm and outcome in opposite directions (last 2 rows, **Table 3**), large bias ensues in the observed $RR_{pos}$ and $RR_{neg}$. The corrections are necessary to remove bias, although note that outcome-differential non-compliance implies that the $RR_{pos}$ corrected to reflect the non-compliance pattern of the screen-arm will not equal the $RR_{pos}$ under perfect compliance. Power for $RR_{pos}$ remains high (84%) and corrected $RR_{neg}$ maintains 5% type-1 error.

## Discussion

We demonstrated that the IE design increases power versus the standard analysis of screening RCTs.  However, the IE design requires assumptions that we relax and investigate sensitivity to.  The IE design can be made more cost-efficient by testing only a stratified sample of control-arm specimens.  Under false-reassurance, the power gain from the IE is even greater than if unintended effects did not exist.   We demonstrate that the IE design is naturally robust to loss-of-signal that non-differential by outcome, and we make it robust to loss-of-signal that is differential by outcome by incorporating "retest-positive" fractions from the screen-arm.  The IE is robust to non-compliance with control-arm specimen collections that is non-differential or differential by arm only, and non-compliance corrections can increase power.  But if non-compliance is differential by both arm and outcome, bias ensues, and corrections must be used.

The IE design reduces costs by requiring fewer participants to achieve 90% power, but increases costs by requiring testing of control-arm specimens. Recruiting a participant for screening RCTs in the US costs roughly $1000-3000, but a screening test (without markup) might cost $100-300 (roughly a factor of 3-10 less). Hence powering an RCT based on the IE design, which would reduce sample-size, should generally have lower total cost than the standard design. In this paper, we further improve cost-efficiency of the IE design by testing only a stratified sample of control-arm specimens. We found that at least half of control-arm specimens do not require testing to maintain nearly all the power. Our simulations did not include sampling strata likely associated with ever-positivity, such cancer-status, age or other risk-factors. Additional strata would reduce further the fraction of control-arm specimens requiring testing. Note that, if power gains from the IE are used to reduce sample-size, then trial investigators must commit to the IE analysis as the primary analysis.

We showed that even if unintended effects occur, the IE analysis still has substantially greater power than the standard analysis. However, "no unintended effects" assumption can be test if RRneg equals 1, which is important to do. If the RRneg indicates substantial false-reassurance, that indicates that screening programs must emphasize to screen-negatives that they must continue to get standard-of-care screens and not change their behavior.

We demonstrate that the IE design is robust to loss-of-signal that is differential only by arm. To account for loss-of-signal that is also differential by outcome, we extended the IE design to also store specimens in the screen-arm to calculate "retest-positive" fractions that are necessary to correct for bias. The only screen-arm stored-specimens that require testing are from those who test ever-positive, to examine whether their stored specimens "retest" ever-positive. For MCD tests, which are rarely positive, this would not appreciably increase testing costs. Because degradation of stored specimens rarely (if ever) results in "gain-of-signal", we assumed it was not possible. If gain-of-signal were possible, then stored screen-arm specimens from never-positives would also have to be tested, which would increase testing costs. Testing only a stratified sample of the never-positives might suffice to correct bias.

The IE design is robust to non-compliance with control-arm specimen collections that is non-differential or differential by arm only, but our corrections can increase statistical power. Non-compliance that is differential by both arm and outcome can cause serious bias in the uncorrected $RR_{pos}$ and $RR_{neg}$. The corrections remove bias and power can be maintained.

**Conclusions**

The IE design is a promising approach to screening RCT design, but requires assumptions that we investigated and relaxed. Because the IE analysis, as we define it, has never been used, we need experience with the design and analysis, which we will obtain with the NCI Vanguard MCD pilot feasibility RCT.[12] If the IE analysis proves feasible and useful, a future large definitive randomized MCD screening trial could be powered for the IE analysis as the primary analysis, which would result in large cost savings versus the standard design and hence make the trial more feasible to conduct.

## References

1. Prorok PC, Andriole GL, Bresalier RS, et al. Design of the Prostate, Lung, Colorectal and Ovarian (PLCO) Cancer Screening Trial. *Control Clin Trials*. 2000;21(6 Suppl):273S-309S.

2. Aberle DR, Berg CD, Black WC, et al. The National Lung Screening Trial: overview and study design. *Radiology*. 2011;258(1):243-253. doi:10.1148/radiol.10091808

3. Menon U, Gentry-Maharaj A, Burnell M, et al. Ovarian cancer population screening and mortality after long-term follow-up in the UK Collaborative Trial of Ovarian Cancer Screening (UKCTOCS): a randomised controlled trial. *The Lancet*. 2021;397(10290):2182-2193. doi:10.1016/S0140-6736(21)00731-5

4. Neal RD, Johnson P, Clarke CA, et al. Cell-Free DNA–Based Multi-Cancer Early Detection Test in an Asymptomatic Screening Population (NHS-Galleri): Design of a Pragmatic, Prospective Randomised Controlled Trial. *Cancers*. 2022;14(19):4818. doi:10.3390/cancers14194818

5. Kisiel JB, Papadopoulos N, Liu MC, Crosby D, Srivastava S, Hawk ET. Multicancer early detection test: Preclinical, translational, and clinical evidence–generation plan and provocative questions. *Cancer*. 2022;128(S4):861-874. doi:10.1002/cncr.33912

6. LeeVan E, Pinsky P. Predictive Performance of Cell-Free Nucleic Acid-Based Multi-Cancer Early Detection Tests: A Systematic Review. *Clinical Chemistry*. Published online October 4, 2023:hvad134. doi:10.1093/clinchem/hvad134

7. Rubinstein WS, Patriotis C, Dickherber A, et al. Cancer screening with multicancer detection tests: A translational science review. *CA A Cancer J Clinicians*. 2024;74(4):368-382. doi:10.3322/caac.21833

8. Katki HA, Prorok PC, Castle PE, Minasian LM, Pinsky PF. Increasing power in screening trials by testing control-arm specimens: application to multicancer detection screening. *JNCI: Journal of the National Cancer Institute*. Published online September 12, 2024:djae218. doi:10.1093/jnci/djae218

9. Weiss NS. The withholding of test results as a means of assessing the effectiveness of treatment in test-positive persons. *Journal of Clinical Epidemiology*. 2013;66(4):355-358. doi:10.1016/j.jclinepi.2012.07.011

10. Hackshaw A, Berg CD. An efficient randomised trial design for multi-cancer screening blood tests: nested enhanced mortality outcomes of screening trial. *The Lancet Oncology*. 2021;22(10):1360-1362. doi:10.1016/S1470-2045(21)00204-7

11. Weiss NS. Randomized trials of multicancer screening tests: augmenting their ability to identify a genuine mortality benefit. *JNCI: Journal of the National Cancer Institute*. 2024;116(7):1005-1007. doi:10.1093/jnci/djae059

12. Minasian LM, Pinsky P, Katki HA, et al. Study design considerations for trials to evaluate multicancer early detection assays for clinical utility. *JNCI: Journal of the National Cancer Institute*. 2023;115(3):250-257. doi:10.1093/jnci/djac218

13. Samimi G, Temkin SM, Weil CJ, et al. Perceptions of Multicancer Detection Tests Among Primary Care Physicians and Laypersons: A Qualitative Study. *Cancer Med*. 2024;13(21):e70281. doi:10.1002/cam4.70281

14. Kitchener HC, Almonte M, Thomson C, et al. HPV testing in combination with liquid-based cytology in primary cervical screening (ARTISTIC): a randomised controlled trial. *Lancet Oncol*. 2009;10(7):672-682. doi:S1470-2045(09)70156-1 [pii] 10.1016/S1470-2045(09)70156-1

15. Baker SG, Lindeman KS. Multiple Discoveries in Causal Inference: LATE for the Party. *CHANCE*. 2024;37(2):21-25. doi:10.1080/09332480.2024.2348956

16. Lachin JM. *Biostatistical Methods: The Assessment of Relative Risks*. John Wiley & Sons; 2009.

**Figure 1.** Schematic of the Intended Effect (IE) analysis. The standard trial analysis table is decomposed into 2 subtables of those who test negative on all screens in each arm, and those who test positive on at least one screen in each arm (note that this requires storing control-arm specimens for future testing). Because those who test negative on all screens never receive treatments, there is no intended effect of screening for the 95,000 screen-negatives and hence $RR_{neg}$ exactly with $p = 1$ exactly. Removing them from the analysis leads to the increases in power shown in the Ever-positive table, which is the Intended Effect analysis.

**Standard trial analysis table**

| | screen | control | |
|---|---|---|---|
| D+ | 900 | 1,000 | 1,900 |
| D- | 49,100 | 49,000 | 98,100 |
| | 50,000 | 50,000 | 100,000 |

- RR = 900/1000 = 0.90
- RD = 2.0% − 1.8% = 0.2%
- p = 0.019
- Power = 65%

- 95% test never test positive on any screen

- 5% test positive on at least 1 screen
- IE analysis: only analyze the ever-positives as they are the only ones affected by screening

**Never-positive table**

| | screen | control | |
|---|---|---|---|
| D+ | 250 | 250 | 500 |
| D- | 47,250 | 47,250 | 94,500 |
| | 47,500 | 47,500 | 95,000 |

- $RR_{neg}$ = 250/250 = 1
- $RD_{neg}$ = 0.53% − 0.53% = 0%

Remove the never-positives because there is no effect of screening in them

- p-value is 10-fold smaller
- Power increases from 65% to 88%

**Ever-positive table**

| | screen | control | |
|---|---|---|---|
| D+ | 650 | 750 | 1,400 |
| D- | 1,850 | 1,750 | 3,600 |
| | 2,500 | 2,500 | 5,000 |

- $RR_{pos}$ = 650/750 = 0.867
- $RD_{pos}$ = 30% − 26% = 4%
- p = 0.0016
- Power = 88%

**Figure 2.** Fixing overall trial RR=0.9 and 2% control-arm outcome prevalence, IE analysis power as a function of $RR_{pos}$ (from 0.5 to 0.9) and ever-positivity (2.5%, 5%, 50%, 90%), for no unintended effects ($RR_{neg}$=1; left), false-reassurance ($RR_{neg}$=1.05; middle), and non-assurance ($RR_{neg}$=0.95; right). The standard analysis has 36% power (solid black line in all 3 plots). Y-axis is on the probit scale. Horizontal dotted lines denote 80% and 90% power.

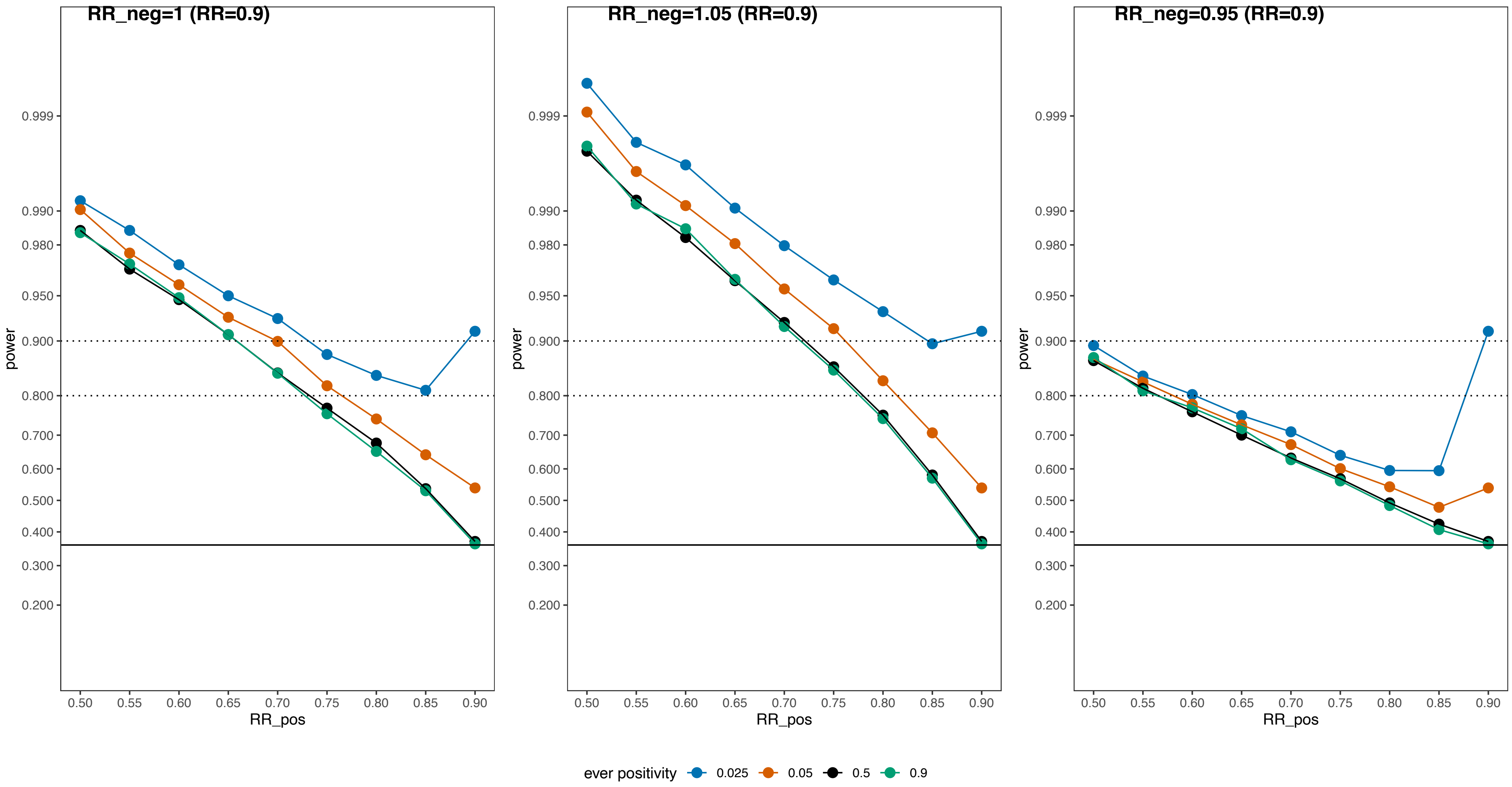

**Table 1A/B.** Power for estimating $RR_{pos}$ by subsampling a fraction of non-events in the control arm for testing all of their stored specimens (all control-arm participants who experience a trial outcome have all their stored specimens tested). Parameters for 3 situations set so that sampling everyone yields 90% power: 5% chance of ever testing positive, 2% event rate, 50% of the total participants are in the control arm, overall $RR = 0.9$, $RR_{neg}$=1 (i.e. no false-reassurance), and no loss-of-signal from testing stored specimens.

**(A)** Sample 100% of events in the control arm for testing of all their stored specimens, and a fraction of the non-events in the control arm for testing of all their stored specimens

| Non-event sampling fraction | Power | | |
|---|---|---|---|
| | n=100,000 $RR_{pos}$=0.867 | n=75,000 $RR_{pos}$ =0.8 | n=50,000 $RR_{pos}$=0.7 |
| 0.1 | 0.54 | 0.64 | 0.67 |
| 0.2 | 0.71 | 0.77 | 0.80 |
| 0.3 | 0.78 | 0.81 | 0.84 |
| 0.4 | 0.81 | 0.85 | 0.85 |
| 0.5 | 0.84 | 0.86 | 0.86 |
| 0.6 | 0.84 | 0.87 | 0.87 |
| 0.7 | 0.86 | 0.88 | 0.88 |
| 0.8 | 0.86 | 0.88 | 0.88 |
| 0.9 | 0.87 | 0.89 | 0.88 |
| 1 | 0.89 | 0.90 | 0.90 |

**(B)** Sample 80% of events in the control arm for testing of all their stored specimens, and a fraction of the non-events in the control arm for testing of all their stored specimens

| Non-event sampling fraction | Power | | |
|---|---|---|---|
| | n=100,000 $RR_{pos}$=0.867 | n=75,000 $RR_{pos}$ =0.8 | n=50,000 $RR_{pos}$=0.7 |
| 0.1 | 0.52 | 0.62 | 0.66 |
| 0.2 | 0.70 | 0.74 | 0.76 |
| 0.3 | 0.77 | 0.79 | 0.81 |
| 0.4 | 0.81 | 0.82 | 0.83 |
| 0.5 | 0.83 | 0.83 | 0.84 |
| 0.6 | 0.85 | 0.85 | 0.85 |
| 0.7 | 0.86 | 0.85 | 0.85 |
| 0.8 | 0.87 | 0.87 | 0.85 |
| 0.9 | 0.88 | 0.87 | 0.85 |
| 1 | 0.87 | 0.87 | 0.86 |

**Table 2A/B/C.** Sensitivity to, and correcting for, loss-of-signal from testing stored specimens that are truly positive. All situations set: total trial n=100,000 (50,000 in each arm), 5% chance of ever testing positive, 2% event rate, overall $RR = 0.9$, $RR_{neg}$=1 (i.e. no false-reassurance), $RR_{pos}$=0.867, and testing 95% of control-arm stored specimens who experience a trial outcome and testing 50% of control-arm stored specimens who do not experience a trial outcome.

**(A)** Non-differential loss-of-signal (i.e. equal loss-of-signal fraction from testing stored specimens from those who experience a trial outcome versus those who do not). $RR_{pos}$ is not biased by non-differential loss-of-signal, with little loss in power at small loss-of-signal fractions of 10%. $RR_{neg}$ is strongly biased even at 10% loss-of-signal with highly inflated type-1 error $\alpha$.

| loss of signal | mean $RR_{pos}$ | mean $RR_{neg}$ | power for $RR_{pos}$ | $\alpha$ for $RR_{neg}$ |
|---|---|---|---|---|
| 50% | 0.867 | 0.41 | 0.66 | 1 |
| 40% | 0.867 | 0.47 | 0.72 | 1 |
| 30% | 0.867 | 0.54 | 0.75 | 1 |
| 20% | 0.867 | 0.63 | 0.80 | 1 |
| 10% | 0.867 | 0.78 | 0.82 | 0.86 |
| 0% | 0.867 | 1.00 | 0.84 | 0.053 |

**(B)** Differential loss-of-signal such that there is more loss-of-signal from testing stored specimens from controls who do not experience a trial outcome versus controls who do experience the trial outcome. With just a 10% loss-of-signal in non-events (and 0% in events), $RR_{pos}$ is underestimated (0.867 vs 0.81), a non-conservative bias. If there is loss-of-signal in events, $RR_{neg}$ is strongly biased even at 10% loss-of-signal in events, with highly inflated type-1 error $\alpha$.

| loss of signal | | | | | |
|---|---|---|---|---|---|
| events | non-events | mean $RR_{pos}$ | mean $RR_{neg}$ | power for $RR_{pos}$ | $\alpha$ for $RR_{neg}$ |
| 50% | 60% | 0.75 | 0.41 | 1 | 1 |
| 40% | 50% | 0.77 | 0.47 | 1 | 1 |
| 30% | 40% | 0.78 | 0.54 | 1 | 1 |
| 20% | 30% | 0.79 | 0.63 | 1 | 1 |
| 10% | 20% | 0.80 | 0.78 | 1 | 0.84 |
| 0% | 10% | 0.81 | 1.01 | 0.99 | 0.048 |
| 0% | 0% | 0.867 | 1.00 | 0.84 | 0.053 |

**(C)** Fixing differential loss-of-signal by storing a portion of specimen in the screen-arm, retesting all stored specimens for those ever screen-positive in the screen-arm, estimating retest-positive fractions among those who experienced the trial event (and did not), and using the retest-positive fractions to correct $RR_{pos}$ and $RR_{neg}$ estimates. $RR_{pos}$ and $RR_{neg}$ estimates are unbiased. Testing for $RR_{neg}$ has proper type-1 error $\alpha$. $RR_{pos}$ estimation loses some power if loss-of-signal is 10% in events and 20% in non-events (power drops from 0.84 to 0.79).

| loss of signal | | | | | |
|---|---|---|---|---|---|
| events | non-events | mean $RR_{pos}$ | mean $RR_{neg}$ | power for $RR_{pos}$ | $\alpha$ for $RR_{neg}$ |
| 50% | 60% | 0.87 | 1.00 | 0.53 | 0.051 |
| 40% | 50% | 0.87 | 1.00 | 0.59 | 0.049 |
| 30% | 40% | 0.87 | 1.00 | 0.67 | 0.051 |
| 20% | 30% | 0.87 | 1.00 | 0.73 | 0.049 |
| 10% | 20% | 0.87 | 1.00 | 0.79 | 0.052 |
| 0% | 10% | 0.87 | 1.00 | 0.82 | 0.048 |
| 0% | 0% | 0.87 | 1.00 | 0.84 | 0.053 |

**Table 3.** Simulating different non-compliance patterns to screening (screen arm) or blood collections (control arm) and calculating the mean $RR_{pos}$ and $RR_{neg}$ and power for $RR_{pos}$, and correcting them for non-compliance. "D+" denotes people who experience the trial primary outcome and "D-" is otherwise. Non-compliance leads to a fraction of true ever-positives being observed as never-positive (first 4 columns). All other simulation parameters set as in **Figure 1**. In simulations, corrected $RR_{neg}$ always maintained 5% type-1 error.

| | non-compliance among ever-positives | | | | Ever-positivity | | | | | | | |
|---|---|---|---|---|---|---|---|---|---|---|---|---|
| | screen arm D+ | screen arm D- | control arm D+ | control arm D- | control arm | screen arm | mean RRpos | mean RRneg | corrected RRpos | corrected RRneg | power for RRpos | power for corrected RRpos |
| perfect compliance | 0% | 0% | 0% | 0% | 5.0% | 5.0% | 0.867 | 1.00 | 0.867 | 1.00 | 88% | 88% |
| non-differential | 30% | 30% | 30% | 30% | 5.0% | 5.0% | 0.867 | 1.01 | 0.867 | 1.01 | 74% | 82% |
| differential by arm | 20% | 20% | 30% | 30% | 5.0% | 5.0% | 0.867 | 1.01 | 0.867 | 1.01 | 77% | 83% |
| differential by outcome | 30% | 10% | 30% | 10% | 4.69% | 4.73% | 0.860 | 1.01 | 0.860 | 1.01 | 77% | 85% |
| differential by outcome & arm | 30% | 15% | 15% | 30% | 5.30% | 4.78% | 0.656 | 0.68 | 0.861 | 1.01 | 100% | 84% |
| | 15% | 30% | 30% | 15% | 4.75% | 5.26% | 1.148 | 1.48 | 0.874 | 1.01 | 75% | 84% |

# Online Supplement

**Table of Contents**

**Review of the Intended Effect design and analysis**

For all details, see Katki et al[8].

The standard trial analysis calculates the relative risk: $RR = P_1(D+) \,/\, P_0(D+)$, where $P_i(D+)$ is the observed rate of a defined trial outcome $D$ (e.g. death from cancer) through a fixed time, for the screening group $i = 1$, and control group $i = 0$. Similarly, the trial risk difference is $RD = P_0(D+) - P_1(D+)$.

For the IE analysis, we store control-group specimens for testing at some future time, which retrospectively reveals which control-group members would have tested positive. Let "$M+$" denote testing positive at any screen ("ever-positive") and "$M-$" denote not testing positive on any screen ("never-positive"). The IE relative risk among ever-positives is

$$RR_{pos} = \frac{P_1(D+|M+)}{P_0(D+|M+)},$$

and similarly the IE risk difference among ever-positives is

$$RD_{pos} = P_0(D+|M+) - P_1(D+|M+).$$

Standard software for analysis of 2x2 tables can calculate $RR_{pos}$, $RD_{pos}$, and their variances. There can be substantially greater power to detect $RR_{pos}$ (or $RD_{pos}$) versus RR.[8] For example in **Figure 1**, power for the standard analysis is 65% but for the IE analysis is 88%. This is because those who test never-positive should not have their outcome risk affected, and hence we expect that $RR_{neg} \approx 1$:

$$RR_{neg} = \frac{P_1(D+|M-)}{P_0(D+|M-)} = 1,$$

or equivalently, $RD_{neg} = P_0(D+|M-) - P_1(D+|M-) = 0$.

The *intended effect* of screening is that all the benefit of screening, as measured by the standard RR, should be concentrated in the $RR_{pos}$. Thus the $RR_{pos}$ will represent a stronger effect size than RR. In **Figure 1**, RR=0.90 but $RR_{pos}$ =0.867. Hence, we expect that there would be more power for $RR_{pos}$ than for RR. In particular, the power gain for the IE analysis (assuming RRneg=1) increases approximately as (1- $RR_{pos}$)/(1-RR) increases, that is, as the mortality reduction in ever-positives (1-RRpos) becomes stronger than that in the overall trial (1-RR).[8]

However, their[8] power calculations are solely for $RR_{neg}$=1. However, there may be unintended effects of screening among the never-positives that affects their outcome risk and $RR_{neg} \neq 1$. This manuscript provides an expression for power allowing any value of $RR_{neg}$.

For example, if those testing negative in the screen-arm are informed of their test result, they may be falsely reassured and forgo standard-of-care screening or change their behavior ("false-reassurance”). In this case, $RR_{neg} > 1$, i.e. screening inadvertently causes harm among screen-arm never-positives. Or $RR_{neg} < 1$ happens if a person does not feel reassured by a negative test and obtains more screening (“non-assurance”). The RR combines effects in ever-positives and never-positives to give a single population-average result, but examining $RR_{pos}$ and $RR_{neg}$ separately provides insight on the sources of screening effects. While focus is primarily on $RR_{pos}$, it is important to calculate $RR_{neg}$ as well. Note that, if a trial protocol allows for blinding of participants to arm, which requires not returning negative results to screen-arm members, there is little reason to believe that those who test-negative change their behavior, and that may ensure that $RR_{neg} \approx 1$.

We believe the IE design satisfies principles of ethical research as long as trial investigators are in true equipoise about whether the screening test can be used to save lives (see[8,13] for more discussion). The IE design ensures that no harm is done to the control-arm, and standard-of-care medical practice is actively encouraged for all study participants. The key question is whether it is appropriate to test the specimens of participants in the control-arm without returning the results. True equipoise suggests that there is no ethical imperative to return test results to control-arm participants, who will be encouraged to undergo standard-of-care screening. For example, some HPV test screening trials have conducted HPV-testing in the control-arm without returning results to participants, and control-arm participants received standard of care screening (cervical cytology).[14] The trial consent form should clearly state that test results would not be returned to control-arm participants. We propose storing specimens and delaying testing towards the end of the study to remove the possibility of acting upon individual control-arm test results.

**Extending the IE analysis to incorporate stratified sampling for testing control-arm specimens**

The IE analysis requires calculating $P_0(D+|M+)$, which is the rate of outcomes among control-arm ever-positives. The most direct way is to test all control-arm specimens, but testing all specimens is costly. Instead, it is more cost-efficient to estimate $P_0(D+|M+)$ by testing only a stratified sample of control-arm specimens. The most important stratum is having the primary outcome (i.e. those who are D+), as those are most likely to test ever-positive. Among everyone else, strata could include any factors that are associated with test-positivity, such as cancer status, age, race, sex, smoking status, standard-of-care screening test results, and other risk factors. Such factors could be identified by analyzing the screen-arm to identify factors associated with screen-arm ever-positivity.

Assume we have S strata denoted s=1 … S each of whom tests a fraction $f_s$ of its members (for simplicity, presume that all specimens for that member would be tested, to definitively establish ever/never-positivity). Denote stratum 1 as the stratum for testing those who are D+ in the control-arm. Then the total number of D+ events among ever-positives is estimated as

$$\widehat{D_1} = \frac{D_1}{f_1},$$

where $D_1$ is the observed number of people both D+ and ever-positive. Note that ideally $f_1=1$, i.e. we can test everyone who is D+ in the control arm, in which case, $D_1$ is the actual number of people who are both D+ and ever-positive in the control-arm. However, it is often the case that there are minor testing or storage issues that preclude testing everyone, and only a high fraction, such as $f_1=0.95$, may be testable in practice.

To calculate the estimated number of ever-positives in the screen arm, assume number $M_s$ tests ever-positive in stratum s, then:

$$\widehat{M} = \widehat{D_1} + \sum_{s=2}^{S} \frac{M_s}{f_s}.$$

Finally we estimate $P_0(D+|M+)$ as $\widehat{D_1}/\widehat{M}$.

An important special case is when D+ represents cancer death. In addition to testing everyone who dies of cancer, it's clearly informative to test everyone with cancer but did not die, and they would form a stratum of their own.

**Derivation of increase in power for $\mathbf{RD_{pos}}$ and $\mathbf{RD}$ in large samples, by calculating the ratio increase in the non-centrality parameter**

Katki et al[8] derived an approximation to the power gain for $RD_{pos}$ over the RD, assuming $RD_{neg}=0$. They approximated power by using the variance calculated under the alternative hypothesis. However, calculating a proper p-value formally requires using the variance as calculated under the null hypothesis.[16] Here we derive the power gain of $RD_{pos}$ over RD for any value of $RD_{neg}$ and we properly use the (pooled) variance as calculated under the null hypothesis.

First, we explain why we focus on the non-centrality parameter $Z$. The non-centrality parameter is the expected value of the Z-statistic under the alternative hypothesis (see page 91 in[16])

$$|Z| = \frac{|\pi_1 - \pi_0|}{\phi},$$

where $\pi_i$ is the probability of the outcome in arm i, and $\phi$ is the pooled variance of $|\pi_1 - \pi_0|$ under the null. The non-centrality parameter (and $\alpha$) suffice to calculate power via the fundamental equation (see page 91 in[16])

$$|Z| = Z_{1-\alpha} - Z_{1-\beta}.$$

The non-centrality parameter for the risk difference RD is a standard calculation. Hence, the ratio of the Z for $RD_{pos}$ and for RD suffices to define the increase in power from $RD_{pos}$ vs RD.

We calculate the ratio of the noncentrality parameters for $RD_{pos}$ and RD:

$$Z_{ratio} = \frac{RD_{pos}}{RD} \cdot \sqrt{\frac{Var(RD)}{Var(RD_{pos})}}.$$

The 2x2 tables for the ever-positives (M+) and the never-positives (M-), by control arm (G=0) and screen arm (G=1) are:

| **M+** | | G=1 | G=0 | | **M−** | | G=1 | G=0 | |
|---|---|---|---|---|---|---|---|---|---|
| | D+ | $a_1$ | $a_0$ | $m_{1+}$ | | D+ | $b_1$ | $b_0$ | $m_{1-}$ |
| | D- | $c_1$ | $c_0$ | $m_{0+}$ | | D- | $d_1$ | $d_0$ | $m_{0-}$ |
| | | $n_{1+}$ | $n_{0+}$ | $N_+$ | | | $n_{1-}$ | $n_{0-}$ | $N_-$ |

The large-sample variance of $RD_{pos}$ under the null is

$$Var(RD_{pos}) = \frac{\pi_{pos}(1-\pi_{pos})N_+}{n_{1+}n_{0+}},$$

where $\pi_{pos} = P(D+|M+) = m_{1+}/N_+$. The large-sample variance of $RD$ is

$$Var(RD) = \frac{\pi(1-\pi)N}{n_1 n_0},$$

where $N = N_+ + N_-, n_1 = n_{1+} + n_{1-}, n_0 = n_{0+} + n_{0-}, \pi = P(D+) = (m_{1+} + m_{1-})/N$. Note that,

$$\frac{\pi(1-\pi)}{\pi_{pos}(1-\pi_{pos})} = \frac{P(D+)P(D-)}{P(D+|M+)(D-|M+)} = \frac{P(M+)^2}{P(M+|D+)P(M+|D-)}.$$

The Z ratio is

$$Z_{ratio} = \frac{RD_{pos}}{RD} \cdot P(M+) \cdot \sqrt{\frac{1}{P(M+|D+)P(M+|D-)}} \cdot \sqrt{\frac{N_+}{n_{1+}} \cdot \frac{n_1}{N} \cdot \frac{n_{0+}}{n_0}}.$$

As the sample size for each arm becomes large, we have

$$\frac{N_+}{n_{1+}} \to \frac{1}{P(G=1|M+)}, \quad \frac{n_1}{N} \to P(G=1), \ and \ \ \frac{n_{0+}}{n_0} \to P(M+|G=0)$$

and thus

$$Z_{ratio} \to \frac{RD_{pos}}{RD} \cdot P(M+) \cdot \sqrt{\frac{1}{P(M+|D+)P(M+|D-)}} \cdot \sqrt{\frac{P(G=1)}{P(G=1|M+)} \cdot P(M+|G=0)}.$$

Under randomization, $M \perp G$ and thus

$$Z_{ratio} \to \frac{RD_{pos}}{RD} \cdot P(M+) \cdot \sqrt{\frac{P(M+)}{P(M+|D+)P(M+|D-)}}.$$

Rewriting the first factor because $RD_{pos} \cdot P(M+) = RD - RD_{neg} \cdot P(M-)$:

$$Z_{ratio} = \left\{1 - \frac{RD_{neg}}{RD} \cdot P(M-)\right\} \cdot \sqrt{\frac{P(M+)}{P(M+|D+)P(M+|D-)}}$$

demonstrates that when $RD_{neg} = 0$, the $Z_{ratio} > 1$ because the radical is always >1 because the numerator is a weighted average of the 2 factors in the denominator. Thus, if $RD_{neg} = 0$, as expected, then the IE analysis has more power than the standard analysis. If there were false-reassurance among never-positives so that $RD_{neg} < 1$, then the $Z_{ratio}$ is even large than under $RD_{neg} = 0$, so power is even more increased. If there were "non-assurance" among never-positives so that $RD_{neg}$ were sufficiently greater than 0, then the $Z_{ratio}$ could be <1, implying reduced power for the IE analysis versus the standard analysis.

Applying Bayes rule to both $P(M+|D-)$ and $P(M+|D+)$:

$$Z_{ratio} = \left\{1 - \frac{RD_{neg}}{RD} \cdot P(M-)\right\} \cdot \sqrt{\frac{P(D-) \cdot P(D+)/P(M+)}{P(D+|M+)\{1 - P(D+|M+)\}}}$$

This expression is amenable to calculating the power gain for IE, by specifying $P_1(D+) = RR \cdot P_0(D+)$ and $P(M+)$. For equal arm sizes, the factors in the denominator of the radical can be written as:

$$P(D+|M+) = 0.5 \cdot \frac{P_0(D+)}{P(M+)} \cdot \left\{(1 + RR_{pos}) \cdot \frac{RR_{neg} - RR}{RR_{neg} - RR_{pos}}\right\}.$$

This factor in the braces contains the quantities related to the IE. The power gain from the IE increases as the factor in the brace decreases. That is, the power gain increases as $RR_{neg}$ increases, and as $RR_{pos}$ decreases relative to RR. Katki et al[8] calculated this factor, under $RR_{neg}$ =1, as (1-RR)/(1- $RR_{pos}$). The extra factor of (1+$RR_{pos}$) in this equation accounts for the non-monotonic power behavior when $RR_{pos}$ is very close to RR at small ever-positivity P(M+) (as shown in **Figure 2** and an example shown in **Figure S1**).

**Simulating screening trials to compare the standard versus the Intended Effect (IE) analysis**

Rather than simulating a screening trial over time, we only simulate the 2 final 2x2 tables of trial outcome by arm (one table each for screen-negative and for screen-positives as shown in **Figure 1**). This is because all stored samples in the control arm would be tested at a single future time after all screening is completed, and the IE analysis would be conducted at that time. We assume there is an overall relative risk (RR) of interest that the trial has been powered to detect at that time.

The simulation fixes the following parameters of the trial:

1. Total trial sample size in both arms (N)
2. Fraction of sample size that is in the control arm
3. Probability of outcome in the control arm
4. Trial relative risk (RR)

These 4 are parameters needed to specify the usual power calculation for the standard analysis. These 4 also specify the 2x2 table for the standard trial analysis (top table in **Figure 1**). Then we also need to specify:

5. Probability of having at least 1 screen-positive (probability of testing ever-positive)
6. $RR_{neg}$
7. $RR_{pos}$

These 3 are needed to define the 2x2 tables for screen-positives (the relative risk among screen-positives $RR_{pos}$) and screen-negatives (the relative risk among screen-positives $RR_{neg}$).

The simulation fixes the sample size in each arm in each simulation (first 2 parameters above define this). The simulation first draws the numbers of people who test positive in each arm from binomial distributions, then draws the number of trial outcomes by the 4 combinations of arm by screening result also from binomial distributions. This populates the 2 2x2 tables for screen-positives and screen-negatives, which are summed to make the overall 2x2 table that is the basis for the standard trial analysis. We do 10,000 simulations for each parameter set.

**Correcting for loss of signal in stored control arm specimens**

For $RD_{pos}$, the only term affected by loss of signal is the control-arm outcome rate among ever-positives, which can be expressed using Bayes Rule as

$$P_0(D+|M+) = \left\{1 + \frac{P_0(D-)}{P_0(D+)} \times \frac{P_0(M+|D-)}{P_0(M+|D+)}\right\}^{-1}.$$

The last ratio is a Bayes Factor (BF), which is the only part affected by ever-positivity and thus what we need to estimate. Denote M as the true ever-positivity and $\widetilde{M}$ as the ever-positivity observed under loss of signal. The numerator and denominator of the BF are calculated via

$$P_0(\widetilde{M}+|D) = P_0(\widetilde{M}+|D,M+)P_0(M+|D) + P_0(\widetilde{M}+|D,M-)P_0(M-|D),$$

and because $P_0(\widetilde{M}+|D,M-) = 0$ (i.e. we assume no gain-of-signal is possible), then

$$P_0(M+|D) = \frac{P_0(\widetilde{M}+|D)}{P_0(\widetilde{M}+|D,M+)}.$$

If loss-of-signal were only differential by arm (i.e. non-differential by outcome, meaning $P_0(\widetilde{M}|D,M) = P_0(\widetilde{M}|M)$), then

$$BF = \frac{P_0(M+|D-)}{P_0\,(M+|D+)} = \frac{P_0(\widetilde{M}+|D-)}{P_0\,(\widetilde{M}+|D+)}.$$

Thus the observed BF equals the true BF, when loss-of-signal is non-differential by outcome. No correction is needed: one can substitute $\widetilde{M}$ for M and directly compute $P_0(D+|M+)$.

For loss-of-signal that is differential by outcome, consider

$$\frac{P_0(\widetilde{M}+|D)}{P_0(\widetilde{M}+|D,M+)} = \frac{P_0(\widetilde{M}+|D,M+)\cdot P_0(M+|D)}{P_0(\widetilde{M}+|D,M+)} = P_0(M+|D).$$

Thus calculating $P_0(M+|D)$ requires $P_0(\widetilde{M}+|D)$, which is observable, but also $P_0(\widetilde{M}+|D,M+)$, which is not observable because true ever-positivity M+ is not observable in the control-arm. However, by testing stored samples in the screen-arm, we can calculate $P_1(\widetilde{M}+|D,M+)$ ("retest-positive fractions") and substitute for $P_0(\widetilde{M}+|D,M+)$. Consider the ratio

$$\frac{P_0(\widetilde{M}+|D-)/P_1(\widetilde{M}+|D-,M+)}{P_0(\widetilde{M}+|D+)/P_1(\widetilde{M}+|D+,M+)} = \frac{P_0(M+|D-)}{P_0\,(M+|D+)} \times \frac{P_0(\widetilde{M}+|D-,M+)/P_1(\widetilde{M}+|D-,M+)}{P_0(\widetilde{M}+|D+,M+)/P_1(\widetilde{M}+|D+,M+)}.$$

The first factor is the true BF that we want to calculate. Under non-differential loss-of-signal, the second factor equals 1, and thus using retest-positive fractions estimates the BF without bias. Under differential loss-of-signal, the second factor still equals 1, but under an assumption. To see this, note that the numerator and denominator of the second factor is

$$\frac{P_0(\widetilde{M}+|D,M+)}{P_1(\widetilde{M}+|D,M+)}=\frac{P_0(D|M+,\widetilde{M}+)\cdot P_0(\widetilde{M}+|M+)/P_0(D|M+)}{P_1(D|M+,\widetilde{M}+)\cdot P_1(\widetilde{M}+|M+)/P_1(D|M+)}.$$

Note that $P_1(D|M+,\widetilde{M}+)=P_1(D|M+)$ (i.e. screen-arm outcomes depend only on true ever-positivity). If also $P_0(D|M+,\widetilde{M}+)=P_0(D|M+)$ (i.e. control-arm outcomes also depend only on true ever-positivity), then all factors on the right-side involving D cancel and thus

$$\frac{P_0(\widetilde{M}+|D,M+)}{P_1(\widetilde{M}+|D,M+)}=\frac{P_0(\widetilde{M}+|M+)}{P_1(\widetilde{M}+|M+)},$$

which is independent of D, and thus the second factor equals 1. Thus

$$\frac{P_0(\widetilde{M}+|D-)/P_1(\widetilde{M}+|D-,M+)}{P_0(\widetilde{M}+|D+)/P_1(\widetilde{M}+|D+,M+)}=\frac{P_0(M+|D-)}{P_0(M+|D+)}$$

estimates the BF without bias, despite substituting screen-arm retest-positivity.

Under differential loss-of-signal, the assumption noted above that is required for lack of bias from substituting screen-arm ever-positivity is

$$P_0(D|M+,\widetilde{M}+)=P_0(D|M+).$$

This says that control-arm outcomes depend on true ever-positivity only. Although reasonable, it could fail in an extreme circumstance. For example, knowing that a person is a true ever-positive, but signal is lost and is observed never-positive, that might mean the person has less chance of being D+ than if you knew that the signal was retained and the person was observed ever-positive.

*Accounting for loss-of-signal to calculate $RD_{neg}$*

For $RD_{neg}$, using Bayes rule, the control-arm outcome rate among never-positives is

$$P_0(D+|M-)=\left\{1+\frac{P_0(D-)}{P_0(D+)}\times\frac{P_0(M-|D-)}{P_0\ (M-|D+)}\right\}^{-1}.$$

The numerator and denominator of the BF are calculated via

$$P_0(\widetilde{M}-|D)=P_0(\widetilde{M}-|D,M+)P_0(M+|D)+\ P_0(\widetilde{M}-|D,M-)P_0(M-|D),$$

and because $P_0(\widetilde{M}- | D, M-) = 1$ (i.e. no gain-of-signal is possible), then

$$P_0(M-|D) = \frac{P_0(\widetilde{M}-|D) - P_0(\widetilde{M}-|D,M+)}{P_0(\widetilde{M}+|D,M+)}.$$

If loss-of-signal were non-differential by outcome (i.e. $P_0(\widetilde{M}|D,M) = P_0(\widetilde{M}|M)$), then

$$BF = \frac{P_0(M-|D-)}{P_0\,(M-|D+)} = \frac{P_0(\widetilde{M}-|D-) - P_0(\widetilde{M}-|M+)}{P_0(\widetilde{M}-|D-) - P_0(\widetilde{M}-|M+)}.$$

Thus the observed BF does not equal the true BF. Thus when loss-of-signal is non-differential by outcome, $RD_{neg}$ is biased and a correction is needed. Following the above derivation for correcting $RD_{pos}$ for differential (or non-differential) loss-of-signal by substituting retest-positive fractions, the BF is unbiasedly estimated by (under the same assumption as for $RD_{pos}$) by

$$\frac{1 - P_0(\widetilde{M}+|D-)/P_1(\widetilde{M}+|D-,M+)}{1 - P_0(\widetilde{M}+|D+)/P_1(\widetilde{M}+|D+,M+)} = \frac{P_0(M-|D-)}{P_0\,(M-|D+)}.$$

Thus $RD_{neg}$ can be unbiasedly estimated by substituting retest-positive fractions.

**Correcting for non-compliance to specimen collections in the control arm**

We focus solely on non-compliance in the control-arm from skipping specimen collections, which is the non-compliance issue specific to the IE design. Those who skipped screens but tested positive at any screen they attended, are known to be ever-positive. The only question is how to handle those with unknown ever-positivity: those who skipped screens and tested negative on every screen they attended.

When you skip a screen, there is no diagnostic intervention or treatments and hence a skipped screen does not affect your outcomes (just like if you had tested negative). Thus, one option is to lump those with unknown ever-positivity into the never-positive table, lumping all participants whom screening does not affect. This makes sense in the screen-arm, because non-compliance to screens is considered part of the effect of screening. That is, if a screen were hard to comply with (e.g. colonoscopy), then that non-compliance plays an unignorable role in the effect of screening. In contrast, some of the non-compliance in the control-arm could happen solely as a part of being in the RCT, which is an artificial effect we want to remove. For example, if control-arm members know they are in the control-arm (i.e. no blinding or imperfect blinding) and know they are not receiving screening, they may choose to non-comply with specimen collections at higher rates than the screen-arm. Thus arm-differential non-compliance is an artifact that might affect the IE analysis that we want to remove.

Thus we examine robustness to lumping those with unknown ever-positivity in the control-arm into the never-positive table. In this framework, non-compliance to specimen collections is mathematically equivalent to a particularly acute loss-of-signal. That is, exactly like this control-arm non-compliance framework, loss-of-signal does not affect the screen-arm, and lumps those true control-arm ever-positives who lost signal into the never-positive table. Then all the math in the **Supplement** section "Correcting for loss of signal in stored control arm specimens" carries over, where now M is the true ever-positivity had the control-arm non-complied in the same way as the screen-arm and $\widetilde{M}$ is the ever-positivity observed under control-arm non-compliance. Hence, just as loss-of-signal that is only differential by arm (i.e. non-differential by outcome) does not cause bias in $RR_{pos}$ (but does cause bias in RRneg), similarly non-compliance that is only differential by arm does not cause bias in RRpos (but does cause bias in $RR_{neg}$). Similarly, just as loss-of-signal that is differential by outcome causes bias in both $RR_{pos}$ and $RR_{neg}$, non-compliance that is differential by outcome also causes bias in both $RR_{pos}$ and $RR_{neg}$.

Correcting for control-arm non-compliance requires separating out everyone with unknown ever-positivity in both arms into a third 2x2 table. **Figure S2** is an example of arm-differential non-compliance: 20% reduction in ever-positivity in the screen-arm and 30% reduction in ever-positivity in the control-arm. To get the tables for non-compliance under **Figure 1** (which is under perfect compliance), in **Figure S2** those tables are calculated by multiplying the control arm cells in the ever/never-positive tables by the arm-specific compliance rate. For example, the ever-positive control-arm D+ cell in **Figure 1** is 750*(1-0.3)=525. Those who are left out of the ever/never-positives tables are placed in the unknown-positive table.

We do not propose to correct for $RR_{pos}$ and $RR_{neg}$ to perfect compliance, as that is unrealistic in screening practice. Screen-arm non-compliance is widely considered a reality of screening to accept. Hence we correct by setting the compliance rate in the control-arm to equal that in the screen arm. In this way, we remove arm-differential non-compliance, and equalize non-compliance between arms. The correction for $RR_{pos}$ requires multiplying the D+ and D- cells in the control-arm ever-positive column by the ratio of compliance rates in the screen-arm to the control-arm, for D+ and D-, respectively. This will force the compliance rate in the control-arm to equal that in the screen arm.

**Figure S2** shows an example of arm-differential non-compliance: 20% reduction in ever-positivity in the screen-arm and 30% reduction in ever-positivity in the control-arm. The observed ever-positive table shows an imbalance in ever-positivity by arm (2000/50000=4% vs. 1750/50000=3.5%). The correction never modifies the numbers in the screen-arm. The control-arm unknown-positivity numbers are corrected by the ratio of the compliance rate in the screen-arm to that in the control-arm, separately for those experiencing the outcome (D+) and not (D-). For example, for control-arm D+, the ratio of compliance rates is (1-180/900)/(1-300/1000) = 0.8/0.7. For control-arm D-, the ratio of compliance rates is (1-9820/49100)/(1-14700/49000) = 0.8/0.7. The ratio of compliance rates is the same for D+ and D- because non-compliance is non-differential by outcome. The corrected ever-positive table has 4% ever-positivity for both arms, and because the correction imposes 20% non-compliance in both arms, the totals in the corrected tables add up to 20% less than the full trial. Most importantly, $RR_{pos}$ and $RR_{neg}$ in the observed and corrected tables are the same and equal to that under perfect compliance. Note that, if we had lumped the unknown-positive table with the observed never-positive table and calculated $RR_{neg}$ (as we proposed to examine robustness), the $RR_{neg}$ would clearly be biased – but the observed ever-positive table is unaffected and $RR_{pos}$ remains unbiased.

**Figure S3** shows an example of non-compliance that is both arm-differential and outcome-differential: (40% in D+ screen-arm, 80% in D+ control-arm, 80% in D- screen-arm, and 40% in D- control-arm). This non-compliance is both strong and is in opposite directions by arm and outcome. This is hopefully unrealistic in practice and chosen solely to make methodological points. The observed $RR_{pos}$=4.1 and $RR_{neg}$ =8.9 are wildly incorrect. The control-arm unknown-positivity numbers are corrected by the ratio of the compliance rate in the screen-arm to that in the control-arm, separately for those experiencing the outcome (D+) and not (D-). For example, for control-arm D+, the ratio of compliance rates is (1-360/900)/(1-800/1000) = 0.6/0.2 = 3. For control-arm D-, the ratio of compliance rates is (1-39280/49100)/(1-19600/49000) = 0.2/0.6 = 1/3. The ratio of compliance rates are reciprocals for D+ and D- because non-compliance is oppositely non-differential by outcome. The corrected ever-positive table has different ever-positivity for each arm (760/50000=1.52% for screen-arm, 800/50000=1.60% for control-arm) because outcome-differential non-compliance does not guarantee that ever-positivity will be balanced across arms (except at the null of RR=1, $RR_{pos}$ =1, and $RR_{neg}$=1). The $RR_{neg}$ in the corrected table happens to equal that under perfect compliance ($RR_{neg}$=1), but that is not guaranteed. Most importantly, the $RR_{pos}$ is not ($RR_{pos}$=0.867 under perfect-compliance, corrected $RR_{pos}$ =0.912). We chose this extreme non-compliance pattern to show that corrected $RR_{pos}$ need not be close to $RR_{pos}$, although in less extreme examples there is little difference. Thus when non-compliance is outcome-differential, the $RR_{pos}$ accounting for screen-arm non-compliance will differ from perfect compliance.

Note that in **Figures S2,S3**, the standard trial analysis table is presumed to already account non-compliance increasing the number of cancer mortality outcomes (D+) in the screen-arm. Thus, although the standard trial analysis table is the same as that in **Figure 1** that assumes perfect compliance, in **Figures S2,S3**, the standard trial analysis table assuming perfect compliance would not be that in **Figure 1**. That is, in **Figures S2,S3**, the standard trial analysis is defined as providing an RR=0.9 in the presence of the chosen level of non-compliance.

A generic assumption required for our corrections is that non-compliance does not depend on true ever/never-positivity given arm and risk-factors and outcome information. If it did, this would be called “non-ignorable non-compliance” or “non-compliance not at random”, and cannot be corrected on the basis of information known to trial investigators. This situation is expected to be unrealistic, as participants should not be able to guess their test result beyond knowing arm and their own risk factors. However, if the screening test were available outside the trial (which we assume is not possible), and those test results were not known by trial investigators, then non-ignorable non-compliance is certainly possible.

**Figure S1.** Example where trial (overall RR=0.8) with larger $RR_{pos}$ =RR=0.8 has more power than $RR_{pos}$ <RR (RRpos=0.7), for very small 2.5% ever-positivity, because $RR_{pos}$ =0.8 has more D+ outcomes than $RR_{pos}$ =0.7 (450 vs 283). This example is the limiting case, fixing RRneg=1, as $RR_{pos} \rightarrow$ RR.

**Standard trial analysis table**

| | screen | control | |
|---|---|---|---|
| D+ | 200 | 250 | 450 |
| D- | 12,300 | 12,250 | 24,550 |
| | 12,500 | 12,500 | 25,000 |

- RR = 200/250 = 0.80
- RD = 2.0% − 1.6% = 0.4%
- Power = 66%
- p = 0.0174

**97.5% test never-positive** (left)

**Never-positive table**

| | screen | control | |
|---|---|---|---|
| D+ | 0 | 0 | **0** |
| D- | 12,187.5 | 12,187.5 | 24,375 |
| | 12,187.5 | 12,187.5 | 24,375 |

- $RR_{neg}$ = 0/0 = 1
- $RD_{neg}$ = 0%
- p = 1

**2.5% test ever-positive**
**$P_{EV\text{-}pos}$=100% of control events in ever-positives**

**Ever-positive table**

| | screen | control | |
|---|---|---|---|
| D+ | 200 | 250 | **450** |
| D- | 112.5 | 62.5 | 175 |
| | 312.5 | 312.5 | 625 |

- **$RR_{pos}$ = 200/250 = 0.8**
- $RD_{pos}$ = 80% − 64% = 16%
- Power = 99.4%
- **p = $8.4 \times 10^{-6}$**
- Fisher p = $1.1 \times 10^{-5}$

**97.5% test never-positive** (right)

**Never-positive table**

| | screen | control | |
|---|---|---|---|
| D+ | 83.33 | 83.33 | **166.67** |
| D- | 12,104.17 | 12,104.17 | 24,108.33 |
| | 12,187.5 | 12,187.5 | 24,375 |

- $RR_{neg}$ = 83.33/83.33 = 1
- $RD_{neg}$ = 0%
- p = 1

**2.5% test ever-positive**
**$P_{EV\text{-}pos}$=166.67/250=66.7% of control events in ever-positives**

**Ever-positive table**

| | screen | control | |
|---|---|---|---|
| D+ | 116.67 | 166.67 | **283.34** |
| D- | 195.83 | 145.83 | 341.66 |
| | 312.5 | 312.5 | 625 |

- **$RR_{pos}$ = 116.67/166.67 = 0.7**
- $RD_{pos}$ = 53.3% − 37.3% = 16%
- Power = 98.5%
- **p = $5.9 \times 10^{-5}$**
- Fisher p = $8.1 \times 10^{-5}$

**Figure S2. Figure 1** with arm-differential non-compliance (20% in screen-arm, 30% in control-arm) showing the Unknown-positivity table. Figures shows the correction to set non-compliance in the control-arm as equal to that in the screen-arm, leading to the corrected ever-positive table and corrected never-positive table.

### Observed never-positive table

**non-compliance: 20% screen-arm, 30% control-arm**

| | screen | control | |
|---|---|---|---|
| D+ | 200 | 175 | |
| D- | 37,800 | 33,075 | |
| | 38,000 | 33,250 | 71,250 |

- 175-200=25 excess deaths
- $RR_{neg}$ = (200/38000)/(175/33250)
  - = 1

### Standard trial analysis table

| | screen | control | |
|---|---|---|---|
| D+ | 900 | 1,000 | 1,900 |
| D- | 49,100 | 49,000 | 98,100 |
| | 50,000 | 50,000 | 100,000 |

- 1000-900=100 prevented deaths
- RR = 900/1000 = 0.90
- p = 0.019, Power = 65%

### Observed ever-positive table

**non-compliance: 20% screen-arm, 30% control-arm**

| | screen | control | |
|---|---|---|---|
| D+ | 520 | 525 | 1,045 |
| D- | 1,480 | 1,225 | 2,705 |
| | 2,000 | 1,750 | 3,750 |

- 525-520=5 prevented deaths (95% fewer)
- $RR_{pos}$ = (520/2000)/(525/1750) = 0.867
- p = 0.0064, Power = 78%

### Unknown positive table

| | screen | control | |
|---|---|---|---|
| D+ | 180 | 300 | |
| D- | 9,820 | 14,700 | |
| | 10,000 | 15,000 | 25,000 |

- 300-180=120 prevented deaths
- $RR_{unk}$ = (180/10000)/(300/15000) = 0.90
- Correction for D+, control
  - Ever-pos: 525 * (1-180/900)/(1-300/1000) = 600
  - Never-pos: 175 * (1-180/900)/(1-300/1000) = 200
- Correction for D-, control
  - Ever-pos: 1225 * (1-9820/49100)/(1-14700/49000) = 1400
  - Never-pos: 33075 * (1-9820/49100)/(1-14700/49000) = 37800

### Corrected never-positive table

| | screen | control | |
|---|---|---|---|
| D+ | 200 | 200 | 400 |
| D- | 37,800 | 37,800 | 75,600 |
| | 38,000 | 38,000 | 76,000 |

- $RR_{pos}$ = 200/200 = 1

### Corrected ever-positive table

| | screen | control | |
|---|---|---|---|
| D+ | 520 | 600 | 1,120 |
| D- | 1,480 | 1,400 | 2,980 |
| | 2,000 | 2,000 | 4,000 |

- $RR_{pos}$ = 520/600 = 0.867
- p=0.0048, Power= 80%

**Figure S3. Figure 1** with arm-differential and outcome-differential non-compliance (40% in D+ screen-arm, 80% in D+ control-arm, 80% in D- screen-arm, and 40% in D- control-arm) showing the Unknown-positive table. Figures shows the correction to set non-compliance in the control-arm as equal to that in the screen-arm, leading to the corrected ever-positive table and corrected never-positive table.

**Standard trial analysis table**

| | screen | control | |
|---|---|---|---|
| D+ | 900 | 1,000 | 1,900 |
| D- | 49,100 | 49,000 | 98,100 |
| | 50,000 | 50,000 | 100,000 |

- 1000-900=100 prevented deaths
- RR = 900/1000 = 0.90

**Observed never-positive table**

| | screen | control | |
|---|---|---|---|
| D+ | 150 | 50 | 200 |
| D- | 9,450 | 28,350 | 37,800 |
| | 9,600 | 28,400 | 38,000 |

- $RR_{neg}$ = (150/9600)/(50/28400) = 8.9

**Observed ever-positive table**

| | screen | control | |
|---|---|---|---|
| D+ | 390 | 150 | 540 |
| D- | 370 | 1,050 | 1,420 |
| | 760 | 1,200 | 1,960 |

- $RR_{pos}$ = (390/760)/(150/1200) = 4.1

**Unknown-positive table**

| | screen | control | |
|---|---|---|---|
| D+ | 360 | 800 | 1,160 |
| D- | 39,280 | 19,600 | 58,880 |
| | 39,640 | 20,400 | 60,040 |

- $RR_{unk}$ = (360/39640)/(800/20400) = 0.23
- Correction for D+, control
  - ever-pos: 150 * (1-360/900)/(1-800/1000) = 450
  - never-pos: 50 * (1-360/900)/(1-800/1000) = 150
- Correction for D-, control
  - ever-pos: 1050 * (1-39280/49100)/(1-19600/49000) = 350
  - never-pos: 28350* (1-39280/49100)/(1-19600/49000) = 9450

- Non-compliance pattern:
  - D+, screen: 40%
  - D+, control:80%
  - D-, screen: 80%
  - D-, control: 40%

**Corrected never-positive table**

| | screen | control | |
|---|---|---|---|
| D+ | 150 | 150 | 300 |
| D- | 9,450 | 9,450 | 18,900 |
| | 9,600 | 9,600 | 19,200 |

- $RR_{neg}$ = 150/150 = 1

**Corrected ever-positive table**

| | screen | control | |
|---|---|---|---|
| D+ | 390 | 450 | 840 |
| D- | 370 | 350 | 720 |
| | 760 | 800 | 1,560 |

- $RR_{pos}$ = (390/760)/(450/800) = 0.912